\documentclass[12pt]{article} \textheight=24 true cm
\usepackage{graphicx}
\usepackage{amsmath}
\usepackage{amssymb}
\begin{document}

 \begin{center}
{\Large\bf Dimension Driven Accelerating Universe
}\\[8 mm]

 S. Chatterjee\footnote{Relativity and Cosmology Research Centre, Jadavpur University,
Kolkata - 700032, India, and also at IGNOU, New Alipore College,
Kolkata  700053, e-mail : chat\_ sujit1@yahoo.com} and
    D. Panigrahi\footnote{Relativity and Cosmology Research Centre,
Jadavpur University, Kolkata - 700032, India, e-mail:
dibyendupanigrahi@yahoo.co.in, Permanent Address : Kandi Raj
College, Kandi, Murshidabad 742137, India\\
Correspondence to : S. Chatterjee},
\end{center}

\begin{abstract}
 The current acceleration of the universe leads us to
investigate higher dimensional gravity theory, which is able to
explain acceleration from a theoretical view point without the
need of introducing  dark energy by hand. We argue that the terms
containing higher dimensional metric coefficients produce an extra
negative pressure that apparently drives an acceleration of the 3D
space, tempting us to suggest that the accelerating universe seems
to  act as a window to the existence of extra spatial dimensions.
Interesting to point out that in this case our cosmology
apparently mimics the well known quintessence scenario fuelled by
a generalised Chaplygin-type of fluid where a smooth transition
from a dust dominated model to a de Sitter like one takes place.
Correspondence to models generated by a tachyonic form of matter
is also briefly discussed.
\end{abstract}

 ~~~Keywords :  cosmology; higher dimensions;
accelerating universe

~~ PACS: 04.20, 04.50 +h

\section{INTRODUCTION}

    ~~~ There are growing evidences today that the current expansion of
 the universe is accelerating. It follows directly from the
 findings of Ia Supernovae  and indirectly from CMBR
 fluctuations. If we put faith in FRW type of models  then General Relativity is unambiguous about
 the need for some sort of dark energy source to explain the
 acceleration, which should behave like a fluid with a large
 negative pressure in the form of a time dependent cosmological
 constant or an evolving scalar field called ~\emph{quintessence}. But
  it is very difficult to construct a theoretical basis for the origin of this exotic matter, which is
 seen precisely at the current epoch when one needs the source for
 cosmic acceleration.

 So there has been a resurgence of interests among relativists,
   field theorists, astrophysicists and people doing astroparticle
   physics both at theoretical and experimental levels to address
   the problems coming out of the recent extra galactic observations
   (for a lucid and fairly exhaustive exposition of some of these
   ideas one is referred to ~\cite{pd} and references therein ) without involving a
   mysterious form of scalar field by hand but looking for
   alternative approaches ~\cite{car} based on sound physical
   principles. Alternatives include, for example,
     modification of the Einstein- Hilbert action through the introduction
      of additional curvature terms, $R^{m}+ R^{n}$ ($ m>0,~ n<0 $ and not
   necessarily integer) in the Lagrangian ~\cite{nkd}. The effective Friedmann
   equations  contain extra terms coming from higher curvatures which may
    be viewed as a fluid, responsible for the current acceleration. However the
    resulting field equations are extremely difficult to solve and
    moreover, the cosmology is mostly unstable against
    perturbations.

   On the other hand, serious attempts are recently being made ~\cite{sc} to
    incorporate the phenomenon of accelerating universe within the framework
    of higher dimensional space time ~\cite{dp} itself without involving any mysterious
scalar field with large negative pressure by hand. The
  realisation that the higher dimensional corrections to the
   Einstein's field equations can be viewed as an effective fluid
   which can emulate the action of the homogeneous part
   of the quintessence field has recently renewed interests in higher
   dimensional model. So in quintessential scenario what we observe as a new
   component of cosmic energy density is, so to say, an effect of higher
   dimensional corrections to the  Einstein-Hilbert action. This
   approach has definite advantage over the standard quintessence
   scenario because we do not need to search for the quintessence
   scalar field and pick them by hand. On the contrary the
   extra fluid responsible for the acceleration is geometrical in
   origin having strong physical foundation and also in line with the spirit
   of general relativity as proposed  by Einstein and others ~\cite{einstein}.
  Assuming a shear free expansion we get a form of solution where
  an additional free parameter appearing in the expression of the
scale factor characterizes the form of the matter field similar to
the well known form of the generalized Chaplygin gas for
quintessential models. The resulting energy momentum tensor
behaves like a mixture of a cosmological constant and a perfect
fluid obeying higher dimensional equation of state. When the
cosmological radius is small the matter field in the form of dust
(for example) predominates giving a decelerating expansion till
the cosmological term takes over effecting a smooth transition to
the current accelerating phase, while in the intermediate stage
our cosmology interpolates between different phases of the
universe. This phenomena has been exhaustively discussed in the
context of quintessence in 4D spacetime. However we are not aware
models of similar kind in higher dimensional spacetime, that too
without assuming by hand any form of an extraneous scalar field
with mysterious properties.

\section{ The FIELD EQUATIONS}

We begin with considering a 5-dimensional line-element

\begin{eqnarray}
  ds^{2} &=&
  dt^{2}-R^{2}\left(\frac{dr^{2}}{1-Kr^{2}}+r^{2}d\theta^{2}+
  r^{2}sin^{2}\theta d\phi^{2}\right) - A^{2}dy^{2}
\end{eqnarray}
where the 3D and extra dimensional scale factors R and A depend on
time only and y is the fifth dimensional co-ordinate and K is the
3D curvature. For our manifold $M^{1}\times S^{3}\times S^{1}$ the
symmetry group of the spatial section is $O(4) \times O(2) $. The
stress tensor whose form will be dictated by Einstein's equations
must have the same invariance leading to the energy momentum
tensor as
\begin{equation}
T_{00}=\rho~,~~T_{ij}= p(t)g_{ij}~,~~ T_{55}=p_{5}(t)g_{55}
\end{equation}
where the rest of the components vanish. Here p is the isotropic
3-pressure and $p_{5}$, that in the fifth dimension. The
independent field equations  for our metric (1) are

\begin{eqnarray}
3\frac{\dot{R^{2}}+ K}{R^{2}} + 3\frac{\dot{R}\dot{A}}{RA} = \rho \\
2\frac{\ddot{R}}{R} + \frac{\dot{R^{2}}+ K}{R^{2}} +
\frac{\ddot{A}}{A}+ 2\frac{\dot{R}\dot{A}}{RA}=  - p    \\
3\frac{\ddot{R}}{R} + 3\frac{\dot{R^{2}}+ K}{R^{2}}= -p_{5}
\end{eqnarray}
Here we have five unknowns ( $A, R, \rho,~ p$ ~and~ $p_{5})$ with
three independent equations and we are at liberty to choose two
connecting equations. We first assume $p = p_{5}$ such that the
field equations ( K = 0) give

\begin{equation}
\ddot{A}+ 2\frac{\dot{R}}{R}\dot{A} - \left(\frac{\ddot{R}}{R} + 2
\frac{\dot{R^{2}}}{R^{2}}\right) A = 0
\end{equation}

 Assuming a form of the deceleration parameter we have rigorously
 solved the above equation, amenable to early decelaration followed by late
 acceleration to account for the current observations. But we defer discussion of
 the above work for a future publication.We see that $A= R$ is a particular
solution of the equation (6). Though simple, in what follows, we
shall presently see that this choice is rich with various
possibilities in interpreting our matter field as also in
comparing the evolution of the universe with a Chaplygin type of
fluid.

As our second assumption we take
\begin{equation}
  A = R = sinh^{n}\omega t
\end{equation}
(where $n$ is an arbitrary constant for the present but we shall
presently see it has strong physical significance) such that

\begin{equation}
\rho = 6n^{2}\omega^{2}+ \frac{6n^{2}\omega^{2}}{sinh^{2}\omega~t}
= \Lambda + \frac{B}{R^\frac{2}{n}}
\end{equation}

\begin{equation}
p = - \Lambda - \frac{2n - 1}{2n} \frac{B}{R^\frac{2}{n}}
\end{equation}
where $A = 6n^{2}\omega^{2}$ and $B = \Lambda
R^{\frac{2}{n}}_{0}$. A little algebra shows that for a perfect
fluid in five dimension with $ p = p_{5}$
 $n=\frac{1}{4}$ corresponds to a stiff fluid
with $\rho \sim \frac{1}{R^{8}}$ and $n=\frac{2}{5}$ to a
radiation dominated phase  with $\rho \sim \frac{1}{R^{5}}$ and
lastly $n=\frac{1}{2}$ to a matter dominated model with $\rho \sim
\frac{1}{R^{4}}$. Thus, interesting to point out that the exponent
$n$  characterizes the nature of the fluid we are dealing with. So
we get the following cases of matter field

a. $(n=\frac{1}{4})$ ( stiff fluid)

\begin{equation}
\rho = \Lambda + \frac{B}{R^{8}} ~~and~~ p = -\Lambda +
\frac{B}{R^{8}}
\end{equation}

b. $(n=\frac{2}{5})$ (radiation)

\begin{equation}
\rho = \Lambda + \frac{B}{R^{5}} ~~and~~ p = -\Lambda +
\frac{B}{4~R^{5}}
\end{equation}

c. $(n=\frac{1}{2})$ ( dust)
\begin{equation}
\rho = \Lambda + \frac{B}{R^{4}} ~~and~~ p = -\Lambda
\end{equation}

We see that for small R the equation in \emph{case a} is
approximated by $\rho = \frac{\Lambda}{R^{8}}$, which corresponds
to a universe dominated by a stiff fluid in 5D spacetime.
Similarly the \emph{case b} and \emph{case c} refer to radiation
dominated and dust dominated universe respectively. On the other
hand for a large value of the cosmological radius we see that the
above equations suggest that $\rho = \Lambda $~~ and~ $p =-
\Lambda$~ which, in turn, corresponds to an empty universe with a
cosmological constant $\Lambda $ (i.e., a de Sitter universe)

Thus equations (10-12) describe the mixture of a cosmological
constant with a type of fluid obeying some equation of state. The
last case known as `stiff fluid' characterized by the equation of
state, $ p = \rho$ is particularly interesting. Note that a
massless scalar field is a particular instance of stiff matter.
Therefore, in a generic situation, our cosmology may be looked
upon as interpolating between different phases of the universe
from a stiff fluid, radiation or dust dominated universe to a de
Sitter one passing through an intermediate phase which is a
mixture just mentioned above. The interesting point, however, is
that such an evolution may be accounted for by using one fluid
only as opposed to the earlier works  ~\cite{gorini} representing
simple two fluid model. Correspondence to models driven by a
generalized Chaplygin type of fluid ~\cite{anjan} described by an
equation of state

\begin{equation}
\rho = \left(\Lambda +
\frac{B}{R^{3(1+\alpha)}}\right)^{\frac{1}{1+\alpha}}
\end{equation}
is only too  apparent although here, as mentioned before we do not
need to hypothesize the existence of a mysterious type of fluid to
explain the observations. Here $\alpha $ is an additional free
parameter to play with to fit the observational data.

Moreover we know that for a sheer-free evolution, if the temporal
dependence of the scale factor is given, one can construct a
potential for a minimally coupled scalar field which would
simulate the evolution as with a perfect fluid. Let us illustrate
the situation in our model. For the Lagrangian

\begin{equation}
L(\phi) = \frac{1}{2}\dot{\phi^{2}} - V(\phi)
\end{equation}
we get the analogous energy density as

\begin{equation}
\rho_{\phi} =\frac{1}{2}\dot{\phi^{2}} + V(\phi) = \Lambda +
\frac{B}{R^{\frac{2}{n}}}
\end{equation}
and the corresponding `pressure' as

\begin{equation}
p_{\phi} =\frac{1}{2}\dot{\phi^{2}} - V(\phi) =- \Lambda -
\frac{B}{R^{\frac{2}{n}}} + \frac{B}{2n~R^{\frac{2}{n}}}
\end{equation}
such that

\begin{equation}
\dot{\phi^{2}} =\frac{B}{2n~R^{\frac{2}{n}}}
\end{equation}
which, in turn, gives via equation (5) for flat 4D space

\begin{equation}
\phi' = \sqrt{\frac{3B}{n}}\frac{1}{R \sqrt{\Lambda
R^{\frac{2}{n}}+B}}
\end{equation}
where $\phi'$ denotes differentiation w. r. t. the scale factor R.
Integrating we get,

\begin{equation}
\phi = \sqrt{\frac{3n}{4}}\ln \frac{\sqrt{B} - \sqrt{\Lambda
R^{\frac{2}{n}}+B}}{\sqrt{B} + \sqrt{\Lambda R^{\frac{2}{n}}+B}}
\end{equation}
Using equation (7) we finally get

\begin{equation}
\phi = \sqrt{3 n}\ln \tanh\frac{\omega t}{2}
\end{equation}
On the other hand simple algebra shows that

\begin{equation}
V ( \phi ) = \Lambda \left( 1 + \frac{1}{2 \sinh^{2} \omega t}
\right)
\end{equation}
For the dust case ( $n = \frac{1}{2}$ )

\begin{equation}
V ( \phi ) = \Lambda \left( 1 + \frac{1}{2 \sinh^{2} \omega t}
\right)
\end{equation}
while for the analogous stiff fluid case ($n = \frac{1}{4}$)~
yields a constant potential $ V(\phi ) = \Lambda = V_{0} $

It may not be out of place to call attention to a quintessential
model driven by a tachyonic scalar field ~\cite{gorini} with a
potential in 4D space time

\begin{equation}
V ( T ) = \frac{\Lambda}{\sin ^{2} \left( \frac{ 3 \sqrt{\Lambda (
1+k )}}{2}\right) T}\sqrt{1 - ( 1+ k) \cos^{2}\left(\frac{ 3
\sqrt{\Lambda ( 1+k )}}{2}\right) T}
\end{equation}
(T is a tachyonic scalar field) giving the cosmological evolution
as

\begin{equation}
R (t) = R_{0} \left( \sinh \frac{3 \sqrt{\Lambda} ( 1+ k )
t}{2}\right)^{\frac{2}{3 ( 1+ k)}}
\end{equation}
It behaves like a two fluid model where one of the fluids is a
cosmological constant while the other obeys a state equation $ p =
k \rho $, $(-1 < k < 1 )$.  Similarity of this evolution with our
model is more than apparent except that we are dealing with a
higher dimensional spacetime. To end the section a final remark
may be in order.From the equations (9-10) we form a sort of
equation of state as

\begin{equation}
p = \frac{1- 2n}{2n}~\rho - \frac{\Lambda}{2n}
\end{equation}
such that the sound speed is given by

\begin{equation}
C_{s}^{2} = \frac{\delta p}{\delta \rho}= \frac{1-2n}{2n}
\end{equation}
which implies that to avoid imaginary value of the speed of sound
$n< \frac{1}{2}$. Evidently in the dust model $(n=\frac{1}{2})$
$C_{s}$ vanishes as expected. This along with the requirement that
$C_{s}$ should never exceed the speed of light further restricts
the range of $n$ as $ \frac{1}{4}< n < \frac{1}{2}$.

\section{Discussion}

In this work we have discussed a 5D homogeneous model with
maximally symmetric 3D space. We have taken only one extra spatial
dimension but we believe most of the findings may be extended if
we take a larger number of extra dimensions. The most important
finding in this work, in our opinion, may be summarised as : we do
not have to hypothesise the existence of an extraneous scalar
field with mysterious properties of matter to achieve an
accelerating universe.  The extra matter field in our model is of
geometrical origin which is, however, not very uncommon in the
literature. Correspondence to curvature quintessence, Wesson's
induced matter theory as also the shadow matter concept of Frolov
etal in the context of brane cosmology may be of some relevance
here. Another point to note is that the whole exercise is based on
assumption of a specific form of the deceleration parameter, which
definitely suffers from the disqualification of a sort of
ad-hocism. But it generates a matter field which is a mixture of
perfect fluid obeying an equation of state as well as a
cosmological constant with either term dominating at different
phases of evolution allowing a smooth transition from a
decelerating to an accelerating model. As a future exercise one
should envisage an additional scenario with other inputs such that
the currently observed acceleration is followed by a decelerating
phase, which finally hits a big brake singularity.

\textbf{Acknowledgment : }

  S.C. wishes to thank TWAS, Trieste for
travel support and ITP (Beijing) for local hospitality where the
work was initiated. The financial support of UGC, New Delhi is
also acknowledged .



\bibliographystyle{aipproc}   


\begin{thebibliography}{15}

\bibitem{pd} T. Padmanabhan  - `Understanding our
Universe : Current status and open issues' , gr- qc / 0503107
\bibitem{car}B. M. N. Carter,  et al, `Type IA supernovae Tests of fractal
buble universe with no cosmic acceleration', astro-ph / 0504192 ;
Zong-Kuan Guo and Y. Z. Zhang , astro-ph / 0506091

\bibitem{nkd} S. Das, N. Banerjee and N. K. Dadhich, `Curvature
driven acceleration: a utopia or a reality ? ', astro - ph /
0505096; Ujjaini Alam, Varun Sahni and A. A. Starobinsky, J. Cosm.
and Astr. Part. Phy.  \textbf{0406}, 008 (2004)
\bibitem{sc}S. Chatterjee, A. Banerjee and Y. Z. Zhang, gr-qc \
0509112;~
 Int. J. Mod. Phy. \textbf{A} 21 4035(2006) ;
 B. Cuadros- Melgar and E. Papantonopoulos, Brazilian J.
 Phys.\textbf{35}, 1117 ( 2005); Li Qiang, Yongge Ma, Moxin Han
 and Dan Yu, Phys. Rev.\textbf{D71}, 061501 (R), 2005.

\bibitem{dp}A. Banerjee, D. Panigrahi  and S. Chatterjee,
 Class. Quantum Grav. \textbf{11}, 1405, (1994).



\bibitem{einstein}A. Einstein, The Meaning of Relativity (
Princeton Univ. Press, Princeton, 1956 ); J. A. Wheeler,
`Einstein's Vision', Springer, Berlin (1968) bibitem{rd}S.

\bibitem{gorini}V. Gorini, A. Kamenshchik, U. Moschella and V.
Pasquier, `Tachyons, scalar fields and cosmology', hep-th /
0311111; J. D. Barrow, Phys. Lett.\textbf{B235}, 40 (1990)
\bibitem{anjan}M. C. Bento, O. Bertolami and A. A. Sen,
Phy. Rev.\textbf{D66}, 043507 (2002).
\end{thebibliography}

\IfFileExists{\jobname.bbl}{}
 {\typeout{}
  \typeout{******************************************}
  \typeout{** Please run "bibtex \jobname" to optain}
  \typeout{** the bibliography and then re-run LaTeX}
  \typeout{** twice to fix the references!}
  \typeout{******************************************}
  \typeout{}
 }



                                             ----------------------------------------------
\end{document}